\documentclass[runningheads,orivec]{llncs}
\usepackage{booktabs}
\usepackage{url}
\usepackage{multirow}
\usepackage{graphicx}
\usepackage{float}
\usepackage[normalem]{ulem}
\useunder{\uline}{\ul}{}
\usepackage{amsfonts}
\usepackage{amsmath}
\DeclareMathOperator*{\argmax}{arg\,max}
\DeclareMathOperator*{\argmin}{arg\,min}
\usepackage[T1]{fontenc}

\usepackage{graphicx,verbatim}

\begin{document}
\title{Test-Time Instance Selection for Improved Whole Slide Image Analysis}

\author{Quoc Anh Nguyen, Sunhong Park, and Jin Tae Kwak}  
\authorrunning{Quoc Anh Nguyen et al.}
\institute{School of Electrical Engineering, Korea University, Seoul 02841, Korea \\
    \email{\{anhnq,sunhongpapa,jkwak\}@korea.ac.kr}}
  
\maketitle             
\begin{abstract}
Whole Slide Image (WSI) analysis has been widely studied for cancer diagnosis. Conventionally, a gigapixel WSI is divided into small patches and processed by Multiple Instance Learning (MIL) models. However, existing MIL models typically process all patches, many of which contain redundant or non-informative tissue patterns. Although recent approaches have focused on instance selection to identify discriminative patches and reduce redundancy, these selection modules still require additional training. In this work, we propose Test-Time Instance Selection (TTIS), a training-free, plug-and-play framework that selects compact yet representative patches during inference. TTIS further incorporates a multi-view ensemble strategy to integrate distinct facets of tissue morphology, enhancing robustness. Importantly, TTIS can be seamlessly integrated into existing MIL models without retraining or architectural changes, enabling flexible deployment. Extensive evaluations across multiple benchmarks demonstrate that our approach improves or matches baseline MIL performance across a range of classification and subtyping tasks. Our implementation code is available at \url{https://github.com/QuIIL/TTIS}


\end{abstract}
\section{Introduction}

Whole slide image (WSI) analysis is a central task in computational pathology for automated diagnosis, cancer subtyping, and outcome prediction~\cite{song2024morphological}. The gigapixel-scale resolution of WSIs, often containing more than 10,000 patches per slide, makes direct end-to-end processing with standard deep neural networks computationally intractable. Multiple instance learning (MIL) has thus become the predominant paradigm for WSI analysis~\cite{dietterich1997solving}, enabling slide-level predictions without requiring exhaustive patch-level annotations. 

Recent advances have improved MIL for WSI analysis along three main directions: (1) self-supervised pathology foundation models that provide robust patch embeddings~\cite{chen2024towards,lu2024visual}, (2), attention-based Transformer architectures for identifying diagnostically relevant patches~\cite{ilse2018attention,lu2021data,shao2021transmil}, and (3) state space models that facilitate efficient long sequence modeling with linear complexity~\cite{fillioux2023structured,yang2024mambamil}. However, conventional MIL pipelines process all patches during both training and inference, many of which are redundant, non-diagnostic, or weakly informative~\cite{kaczmarzyk2024explainable,song2024morphological}. This redundancy not only increases aggregation cost, but can also dilute sparse discriminative information with noise and repetitive patterns.

To address this, recent works have proposed patch selection strategies to mitigate redundancy. For example, DC-WSI~\cite{liang2024enhancing} and MiCo~\cite{li2025mico} use \texttt{K-Means}~\cite{macqueen1967multivariate} and learnable context-aware clustering, respectively, while PAMoE~\cite{wu2025learning} and PANTHER~\cite{song2024morphological} employ prototypes derived from foundation models or global cluster centroids.
While effective, these methods typically rely on additional training objectives, trainable selection modules, or access to training data, which limit applicability to deployed models.
This motivates a practical question: \emph{can we improve predictive performance by mitigating patch redundancy at inference time, without retraining?} To the best of our knowledge, training-free, test-time mechanisms for MIL-based WSI inference remain underexplored.

To this end, we introduce a plug-and-play, training-free test-time instance selection (\texttt{TTIS}) framework for WSI analysis. The key insight is that a compact yet diverse subset of instances/patches can preserve essential information needed for slide-level prediction while suppressing redundancy and noise, and that such a subset can be retrieved directly at inference time using only patch embeddings from the input slide, i.e., without additional training or modification of MIL architectures. 
Given a trained MIL model, \texttt{TTIS} adopts a two-stage clustering approach to capture both spatial and semantic heterogeneity, followed by a stratified sampling to select representative patches while minimizing redundancy. To further improve robustness, we draw multiple patch subsets and utilize an ensemble strategy that combines the resulting slide-level embeddings into a final slide-level representation. We validate \texttt{TTIS} by integrating it with several widely used MIL models. Experimental results across multiple cancer classification and subtyping datasets demonstrate consistent performance improvements. Our implementation will be made publicly available.

\section{Methodology}
\subsection{Problem Formulation}
\begin{figure}
    \centering
    \includegraphics[width=1.0\textwidth]{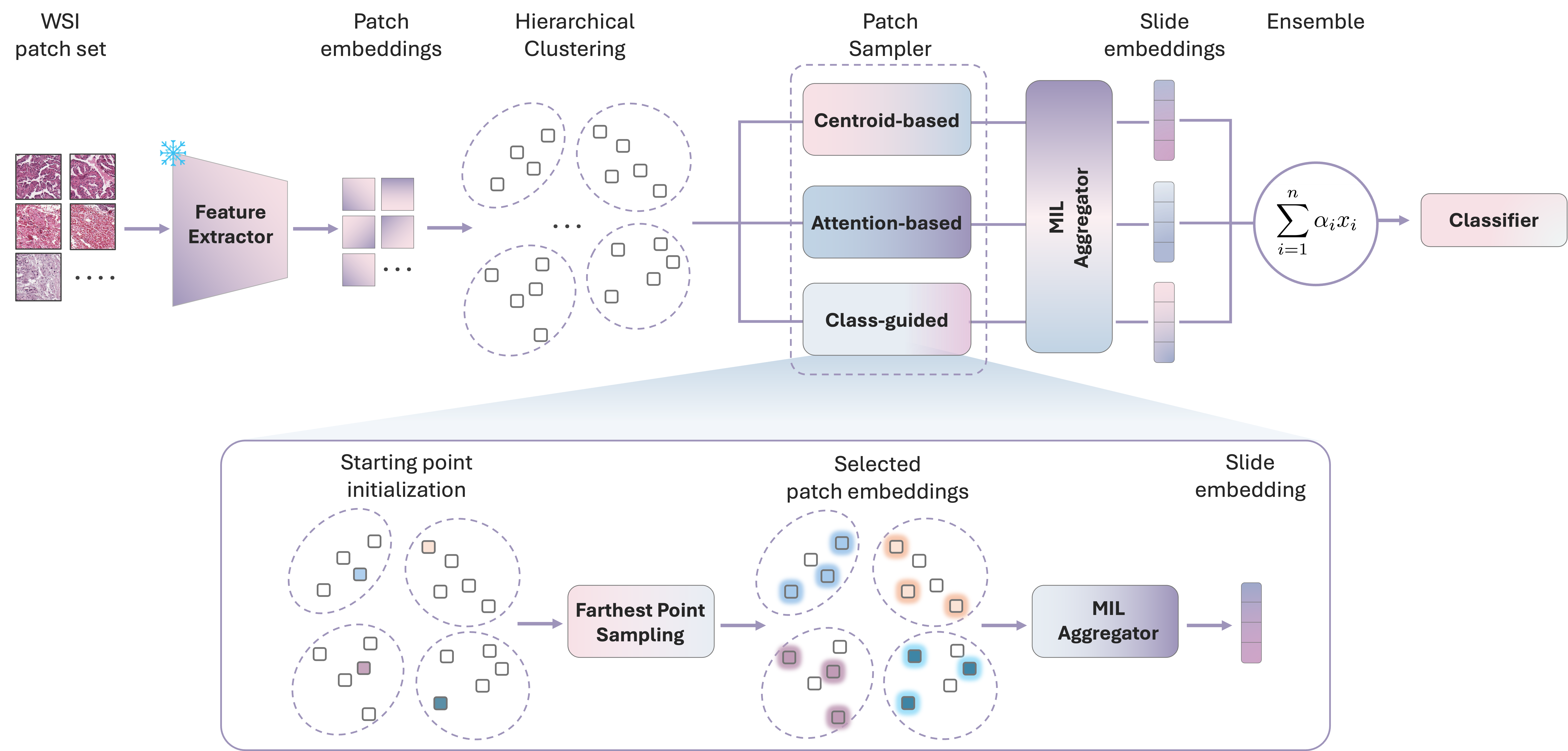}
    \caption{Overview of the \texttt{TTIS} framework.}
    \label{fig:overview}
\end{figure}

Let $I$ denote a WSI and $Y$ its corresponding slide-level label. First, $I$ is divided into a bag of $N$ patches $\mathcal{X} = \{x_i\}_{i=1}^{N}$, where each patch $x_i$ represents a tissue region cropped from the WSI. Then, a feature extractor $\texttt{F}(\cdot)$ transforms each patch into an embedding space, yielding the set of patch embeddings $\mathcal{H} = \texttt{F}(\mathcal{X)} = \{h_i\}_{i=1}^{N}$ where $h_i \in \mathbb{R}^{D_{\text{h}}}$ and $D_{h}$ is its dimension. 
An MIL module $\texttt{MIL}(\cdot)$ aggregates $\mathcal{H}$ into a slide-level embedding $e = \texttt{MIL}(\mathcal{H}) \in \mathbb{R}^{D_{\text{h}}}$, which is used to produce the slide label. In practice, $N$ can be very large, and many patches may be redundant or non-informative. Motivated by recent works on MIL that pinpoint the uneven distribution of patches, our objective is to construct an efficient and effective \texttt{TTIS} framework that identifies a compact, informative subset at inference time:
\begin{equation}
    \mathcal{H}' = \texttt{TTIS}(\mathcal{H}) \quad \text{such that} \quad \mathcal{H}' \subset \mathcal{H}, \quad |\mathcal{H}'| = M \ll N
\end{equation}
where $M$ is the subset size. We then compute $e' = \texttt{MIL}(\mathcal{H}') \in \mathbb{R}^{D_{\text{h}}}$, that maintains or even improves the slide-level prediction compared to using the exhaustive set $\mathcal{H}$. 
An overview of our framework is illustrated in Fig.~\ref{fig:overview}.

\subsection{Test-Time Instance Selection Framework} \label{ttis}
\texttt{TTIS} operates in three steps: (1) hierarchical spatio-semantic clustering to partition patches into groups capturing both spatial organization and morphological similarity, (2) stratified deterministic sampling to select representative patches while maximizing diversity, and (3) ensemble aggregation across multiple sampling configurations to improve and stabilize predictions. 

\subsection{Hierarchical Spatio-Semantic Clustering} \label{clustering}
\texttt{TTIS} employs a hierarchical clustering strategy to partition the patch embeddings $\mathcal{H}$ into groups that preserve both spatial coverage and semantic diversity. It utilizes a two-stage \texttt{K-Means++}~\cite{arthur2006k} clustering, which provides robust centroid initialization and faster convergence compared to standard \texttt{K-Means}. Specifically, we first cluster patches based on their 2D spatial coordinates into $K_1$ clusters, capturing broad spatial organization. Within each spatial cluster, patches are further sub-clustered based on their feature embeddings into $K_2$ sub-clusters, representing local semantic heterogeneity~\cite{Bui_FALFormer_MICCAI2024}. This hierarchical process yields a total of $K_1 \times K_2$ sub-clusters. To ensure balanced granularity at both hierarchical levels, we set $K_1 = K_2 = K$, resulting in $K^2$ sub-clusters in total. 

\subsection{Stratified Sampling}
Given the sub-clusters, \texttt{TTIS} conducts stratified sampling to select a compact yet diverse set of representative patches. We adopt Farthest Point Sampling (\texttt{FPS})~\cite{eldar1997farthest} as the core sampling method and use diverse initialization strategies to enhance representativeness.

\subsubsection{Farthest Point Sampling}
\texttt{FPS} is a greedy downsampling algorithm widely applied in 3D point cloud analysis~\cite{qi2017pointnet++,wu2023attention}. For each sub-cluster $\mathcal{S}$, \texttt{FPS} constructs a subset $\mathcal{Q} \subset \mathcal{S}$ by iteratively selecting the patch that is farthest from the current selected set. Formally, at iteration $t$ with current set $\mathcal{Q}_{t-1}$, the next patch is chosen as: 

\begin{equation}
q_{t} = \argmax_{s\in\mathcal{S} \setminus \mathcal{Q}_{t-1}}\ \min_{q\in\mathcal{Q}_{t-1}} ||h_s - h_q||_2.
\end{equation}
This procedure typically starts from an initial patch $q_{1}$, which is chosen randomly or by a deterministic rule, and repeats the above step until the target subset size is reached. This encourages broad coverage of the sub-cluster $\mathcal{S}$, while preventing over-representation of densely clustered regions.

\subsubsection{Starting Point Selection}
\texttt{FPS} is highly sensitive to starting point selection, which can yield inconsistent subsets across runs. To ensure deterministic behavior and enhance subset representativeness, we explore three deterministic initialization criteria.
First, \textit{centroid-based} initialization selects the patch closest to each sub-cluster centroid $\mathbf{c}$ as the \texttt{FPS} starting point: $s_{\text{ctr}} = \argmin_{s \in \mathcal{S}} \lVert h_s - \mathbf{c} \rVert_2$.
Second, \textit{attention-based} initialization leverages attention scores generated by attention-based MIL models~\cite{ilse2018attention,lu2021data,shao2021transmil}. Within each sub-cluster $\mathcal{S}$, we select the patch with the highest attention score as the \texttt{FPS} starting point: $s_{\text{attn}} = \argmax_{s \in \mathcal{S}} a_s$,where ${a_s}$ denotes the attention score assigned to patch $s$.
For MIL models that do not output explicit attention scores, we derive pseudo-attention scores using softmax-normalized $L_2$ norms of the patch embeddings: $a_s = \frac{\exp(\lVert h_{s} \rVert_2)}{\sum_{j \in \mathcal{S}} \exp(\lVert h_{j} \rVert_2)}$.
Third, \textit{class-guided} initialization employs a pretrained pathology foundation model (e.g., TITAN~\cite{ding2025multimodal} and MUSK~\cite{xiang2025vision}) to identify the most diagnostically relevant patch as the starting point for each class. Given $C$ task-specific diagnostic categories, we define $K$ text prompts for each class following the standard practice of~\cite{lu2024visual}. A text encoder produces prompt embeddings $\mathcal{E} = \{e_k^{(c)} \mid c \in \{1, \ldots, C\}, k \in \{1, \ldots, K\}\}$. Averaging prompt embeddings per class yields the class prototype embeddings $\bar{\mathcal{E}} = \{\bar{e}^{(c)}\}_{c=1}^{C}$ where $\bar{e}^{(c)} \in \mathbb{R}^{D_{\text{t}}}$ and $D_{\text{t}}$ is the prompt text dimension.
For each patch $s \in \mathcal{S}$, we calculate the cosine similarity between its embedding $h_{s}$ and each class prototype embedding:
\begin{equation}
\text{sim}_s^{(c)} = \frac{h_{s} \cdot \bar{e}^{(c)}}{\lVert h_{s}\rVert_2 \cdot \lVert\bar{e}^{(c)}\rVert_2}.
\end{equation}
The patch with the highest similarity to each class $c$ becomes the class-specific starting point: $s_{\text{cls}}^{(c)} = \argmax_{s \in \mathcal{S}} \text{sim}_s^{(c)}$. This yields $C$ class-specific starting points per sub-cluster for $\texttt{FPS}$. 

\subsection{Multi-view Test-time Ensemble}
Each initialization criterion emphasizes a unique facet of tissue representations, spanning semantic representativeness (centroid), model-specific importance (attention), and diagnostic specificity (class-guided). Relying on a single criterion may discard complementary information. To maximize robustness and predictive consistency, we propose a multi-view test-time ensemble that produces a slide-level embedding for each criteria and aggregates these embeddings using softmax-normalized weights. The aggregated embedding is then fed into the classifier to produce the final slide-level prediction.

\section{Experiments}
\subsection{Datasets}

We evaluate \texttt{TTIS} on five publicly available datasets. We use CAMELYON16 dataset~\cite{ehteshami2017diagnostic}, comprising 399 WSIs, for binary tumor vs. non-tumor classification, and four TCGA datasets~\cite{weinstein2013cancer} for cancer subtyping: (1) Breast (TCGA-BRCA), containing 875 WSIs of invasive ductal and invasive lobular carcinoma, (2) Esophagus (TCGA-ESCA), containing 954 WSIs of adenocarcinoma and squamous cell carcinoma, (3) Kidney (TCGA-RCC), containing 905 WSIs of papillary, chromophobe, and clear cell carcinoma, and (4) Lung (TCGA-NSCLC), containing 167 WSIs of adenocarcinoma and squamous cell carcinoma.

\subsection{Implementation}
We integrate \texttt{TTIS} with six MIL models: ABMIL~\cite{ilse2018attention}, TransMIL~\cite{shao2021transmil}, CLAM-SB~\cite{lu2021data}, WiKG~\cite{li2024dynamic}, S4MIL~\cite{fillioux2023structured}, and R$^2$T-MIL~\cite{tang2024feature}. All WSIs are processed at $20\times$ magnification using non-overlapping $256\times256$-pixel patches. Patch features are extracted using UNI~\cite{chen2024towards}, and the MUSK~\cite{xiang2025vision} text encoder is employed for class-guided starting point initialization. We set the sampling ratio to $r=30\%$. For clustering, \texttt{K-Means++} centroid initialization is performed 10 times, and the number of clusters $K$ is set to 16 for all datasets except TCGA-ESCA, where $K=2$ due to smaller average patch counts per WSI. To ensure validity, all experiments are conducted with 3-fold cross-validation under a fixed random seed and implemented on a single NVIDIA RTX A6000 GPU.

\begin{table}[ht]
\centering
\caption{Results of WSI classification ($mean_{std}$). Random sampling results are averaged over 30 runs. The best and second-best results are highlighted in \textbf{bold} and \underline{underlined}, respectively.}
\label{tab:results-main}
\setlength{\tabcolsep}{4pt}
\renewcommand{\arraystretch}{1}
\resizebox{0.9\textwidth}{!}{%
\begin{tabular}{l c c|c c c|c c c c}
\toprule
Dataset & Instance Selection & Model & ACC & AUC & F1 & Model & ACC & AUC & F1 \\
\midrule
 \multirow{9}{*}{CAMELYON16}&None & \multirow{3}{*}{ABMIL}& $\underline{96.64_{1.18}}$ & $\textbf{0.994}_{\textbf{0.002}}$ & $\underline{0.964_{0.013}}$  & \multirow{3}{*}{TransMIL}& $\underline{91.99_{1.18}}$ & $\underline{0.965_{0.013}}$ &$\underline{0.912_{0.014}}$  \\
  &Random& & $94.06_{1.04}$ & $0.984_{0.008}$ & $0.935_{0.012}$  & & $89.59_{0.65}$ & $0.953_{0.015}$ &$0.882_{0.008}$  \\
  &\texttt{TTIS}& & $\textbf{97.16}_{\textbf{0.44}}$ & $\underline{0.992_{0.001}}$ & $\textbf{0.969}_{\textbf{0.005}}$  & & $\textbf{93.28}_{\textbf{2.37}}$ & $\textbf{0.974}_{\textbf{0.007}}$ &$\textbf{0.925}_{\textbf{0.028}}$  \\
  \cmidrule(lr){2-10}
 &None& \multirow{3}{*}{CLAM-SB}& $\underline{92.76_{0.44}}$ & $\underline{0.984_{0.002}}$ & $\underline{0.920_{0.005}}$  & \multirow{3}{*}{WiKG}& $\underline{86.30_{9.82}}$ & $\textbf{0.896}_{\textbf{0.099}}$ &$\underline{0.851_{0.107}}$  \\
  &Random & & $91.55_{0.56}$ & $0.964_{0.005}$ & $0.906_{0.007}$  & & $83.77_{6.85}$ & $0.868_{0.098}$ &$0.812_{0.082}$  \\
  &\texttt{TTIS} & & $\textbf{96.38}_{\textbf{1.18}}$ & $\textbf{0.986}_{\textbf{0.002}}$ & $\textbf{0.961}_{\textbf{0.013}}$  & & $\textbf{87.85}_{\textbf{7.65}}$ & $\underline{0.886_{0.110}}$ &$\textbf{0.864}_{\textbf{0.088}}$  \\
 \cmidrule(lr){2-10}
 &None& \multirow{3}{*}{S4MIL}& $\underline{88.89_{7.53}}$ & $\underline{0.936_{0.058}}$ & $\underline{0.870_{0.095}}$  & \multirow{3}{*}{R$^2$T-MIL}& $\underline{93.28_{3.82}}$ & $\textbf{0.987}_{\textbf{0.009}}$ &$\underline{0.927_{0.042}}$  \\
  &Random & & $84.70_{4.10}$ & $0.920_{0.053}$ & $0.818_{0.056}$  & & $90.98_{3.89}$ & $\underline{0.970_{0.014}}$ &$0.898_{0.045}$  \\
  &\texttt{TTIS} & & $\textbf{89.41}_{\textbf{6.22}}$ & $\textbf{0.942}_{\textbf{0.050}}$ & $\textbf{0.879}_{\textbf{0.075}}$  & & $\textbf{94.06}_{\textbf{3.13}}$ & $\textbf{0.987}_{\textbf{0.008}}$ &$\textbf{0.935}_{\textbf{0.035}}$  \\
  \midrule
\multirow{9}{*}{TCGA-BRCA}
& None   & \multirow{3}{*}{ABMIL}    & $89.85_{3.65}$ & $\underline{0.968_{0.021}}$ & $0.825_{0.076}$
         & \multirow{3}{*}{TransMIL} & $\underline{90.33_{0.84}}$ & $\underline{0.959_{0.018}}$ & $\underline{0.823_{0.033}}$ \\
& Random &                           & $\underline{90.15_{3.59}}$ & $\underline{0.968_{0.021}}$ & $\underline{0.830_{0.073}}$
         &                          & $90.29_{2.90}$ & $\textbf{0.961}_{\textbf{0.017}}$ & $0.816_{0.072}$ \\
& \texttt{TTIS}   &                           & $\textbf{91.94}_{\textbf{4.32}}$ & $\underline{0.968_{0.019}}$ & $\textbf{0.854}_{\textbf{0.088}}$
         &                          & $\textbf{90.88}_{\textbf{2.55}}$ & $0.958_{0.022}$ & $\textbf{0.831}_{\textbf{0.061}}$ \\
\cmidrule(lr){2-10}
& None   & \multirow{3}{*}{CLAM-SB}  & $\textbf{92.82}_{\textbf{3.33}}$ & $\textbf{0.970}_{\textbf{0.017}}$ & $\textbf{0.875}_{\textbf{0.059}}$
         & \multirow{3}{*}{WiKG}     & $90.27_{4.13}$ & $0.928_{0.067}$ & $0.817_{0.104}$ \\
& Random &                           & $\underline{92.70_{3.30}}$ & $\underline{0.969_{0.017}}$ & $\underline{0.873_{0.058}}$
         &                          & $\underline{90.28_{4.65}}$ & $\underline{0.932_{0.062}}$ & $\underline{0.817_{0.107}}$ \\
& \texttt{TTIS}   &                           & $91.97_{3.39}$ & $0.968_{0.017}$ & $0.847_{0.070}$
         &                          & $\textbf{91.72}_{\textbf{3.54}}$ & $\textbf{0.933}_{\textbf{0.062}}$ & $\textbf{0.853}_{\textbf{0.075}}$ \\
\cmidrule(lr){2-10}
& None   & \multirow{3}{*}{S4MIL}    & $91.59_{0.82}$ & $\textbf{0.965}_{\textbf{0.012}}$ & $\underline{0.854_{0.019}}$
         & \multirow{3}{*}{R$^2$T-MIL} & $\underline{90.72_{2.19}}$ & $\underline{0.957_{0.024}}$ & $\underline{0.827_{0.066}}$ \\
& Random &                           & $\underline{91.76_{0.29}}$ & $0.943_{0.044}$ & $\underline{0.854_{0.005}}$
         &                            & $90.80_{2.42}$ & $\textbf{0.958}_{\textbf{0.023}}$ & $\underline{0.827_{0.072}}$ \\
& \texttt{TTIS}   &                           & $\textbf{92.00}_{\textbf{0.95}}$ & $\underline{0.944_{0.039}}$ & $\textbf{0.855}_{\textbf{0.020}}$
         &                            & $\textbf{91.13}_{\textbf{2.76}}$ & $\underline{0.957_{0.025}}$ & $\textbf{0.834}_{\textbf{0.072}}$ \\

\midrule

\multirow{9}{*}{TCGA-ESCA}
& None   & \multirow{3}{*}{ABMIL}    & $\underline{92.19_{4.62}}$ & $\underline{0.970_{0.019}}$ & $\underline{0.920_{0.047}}$
         & \multirow{3}{*}{TransMIL} & $\underline{90.40_{3.81}}$ & $\textbf{0.963}_{\textbf{0.027}}$ & $\underline{0.902_{0.039}}$ \\
& Random &                           & $92.13_{4.04}$ & $\underline{0.970_{0.020}}$ & $\underline{0.920_{0.041}}$
         &                          & $90.09_{5.27}$ & $\underline{0.962_{0.025}}$ & $0.898_{0.055}$ \\
& \texttt{TTIS}   &                           & $\textbf{94.59}_{\textbf{3.19}}$ & $\textbf{0.972}_{\textbf{0.023}}$ & $\textbf{0.945}_{\textbf{0.032}}$
         &                          & $\textbf{91.59}_{\textbf{5.22}}$ & $\textbf{0.963}_{\textbf{0.026}}$ & $\textbf{0.914}_{\textbf{0.054}}$ \\
\cmidrule(lr){2-10}
& None   & \multirow{3}{*}{CLAM-SB}  & $\underline{91.00_{4.81}}$ & $\textbf{0.974}_{\textbf{0.018}}$ & $\underline{0.908_{0.050}}$
         & \multirow{3}{*}{WiKG}     & $\textbf{92.20}_{\textbf{2.83}}$ & $\underline{0.967_{0.035}}$ & $\textbf{0.921}_{\textbf{0.029}}$ \\
& Random &                           & $90.45_{5.38}$ & $\textbf{0.974}_{\textbf{0.020}}$ & $0.902_{0.056}$
         &                          & $\underline{91.11_{3.62}}$ & $0.965_{0.039}$ & $\underline{0.910_{0.037}}$ \\
& \texttt{TTIS}   &                           & $\textbf{91.60}_{\textbf{2.81}}$ & $\underline{0.973_{0.023}}$ & $\textbf{0.915}_{\textbf{0.029}}$
         &                          & $90.99_{3.69}$ & $\textbf{0.968}_{\textbf{0.034}}$ & $0.909_{0.037}$ \\
\cmidrule(lr){2-10}
& None   & \multirow{3}{*}{S4MIL}    & $\underline{89.77_{8.21}}$ & $\textbf{0.961}_{\textbf{0.035}}$ & $\underline{0.895_{0.086}}$
         & \multirow{3}{*}{R$^2$T-MIL} & $\underline{87.39_{4.91}}$ & $\textbf{0.943}_{\textbf{0.031}}$ & $\underline{0.870_{0.053}}$ \\
& Random &                           & $87.98_{9.03}$ & $0.950_{0.034}$ & $0.875_{0.095}$
         &                            & $86.91_{5.36}$ & $\underline{0.942_{0.032}}$ & $0.865_{0.058}$ \\
& \texttt{TTIS}   &                           & $\textbf{90.40}_{\textbf{5.22}}$ & $\underline{0.955_{0.035}}$ & $\textbf{0.902}_{\textbf{0.053}}$
         &                            & $\textbf{89.19}_{\textbf{4.89}}$ & $0.937_{0.035}$ & $\textbf{0.890}_{\textbf{0.050}}$ \\

\midrule

\multirow{9}{*}{TCGA-RCC}
& None   & \multirow{3}{*}{ABMIL}    & $\textbf{96.16}_{\textbf{2.39}}$ & $\textbf{0.995}_{\textbf{0.005}}$ & $\textbf{0.956}_{\textbf{0.022}}$
         & \multirow{3}{*}{TransMIL} & $95.00_{1.74}$ & $\textbf{0.997}_{\textbf{0.002}}$ & $0.940_{0.023}$ \\
& Random &                           & $\underline{95.93_{2.68}}$ & $\textbf{0.995}_{\textbf{0.005}}$ & $\underline{0.952_{0.026}}$
         &                          & $\underline{95.58_{1.83}}$ & $\underline{0.996_{0.003}}$ & $\underline{0.943_{0.025}}$ \\
& \texttt{TTIS}   &                           & $95.78_{2.89}$ & $\textbf{0.995}_{\textbf{0.004}}$ & $0.951_{0.030}$
         &                          & $\textbf{95.77}_{\textbf{1.74}}$ & $\underline{0.996_{0.004}}$ & $\textbf{0.946}_{\textbf{0.025}}$ \\
\cmidrule(lr){2-10}
& None   & \multirow{3}{*}{CLAM-SB}  & $\textbf{93.86}_{\textbf{3.50}}$ & $\textbf{0.995}_{\textbf{0.004}}$ & $\textbf{0.923}_{\textbf{0.043}}$
         & \multirow{3}{*}{WiKG}     & $\textbf{95.77}_{\textbf{2.38}}$ & $\textbf{0.993}_{\textbf{0.006}}$ & $\textbf{0.946}_{\textbf{0.017}}$ \\
& Random &                           & $\textbf{93.86}_{\textbf{3.50}}$ & $\textbf{0.995}_{\textbf{0.005}}$ & $\textbf{0.923}_{\textbf{0.043}}$
         &                          & $94.70_{2.32}$ & $\textbf{0.993}_{\textbf{0.006}}$ & $0.936_{0.017}$ \\
& \texttt{TTIS}   &                           & $\textbf{93.86}_{\textbf{3.50}}$ & $\textbf{0.995}_{\textbf{0.005}}$ & $\textbf{0.923}_{\textbf{0.043}}$
         &                          & $\underline{95.39_{2.29}}$ & $\textbf{0.993}_{\textbf{0.007}}$ & $\underline{0.943_{0.017}}$ \\
\cmidrule(lr){2-10}
& None   & \multirow{3}{*}{S4MIL}    & $92.31_{0.62}$ & $\textbf{0.996}_{\textbf{0.001}}$ & $0.902_{0.003}$
         & \multirow{3}{*}{R$^2$T-MIL} & $\textbf{94.25}_{\textbf{5.01}}$ & $\textbf{0.994}_{\textbf{0.006}}$ & $\textbf{0.939}_{\textbf{0.053}}$ \\
& Random &                           & $\underline{92.69_{0.92}}$ & $\underline{0.995_{0.001}}$ & $\underline{0.912_{0.010}}$
         &                            & $\underline{94.05_{5.06}}$ & $\underline{0.993_{0.007}}$ & $0.936_{0.054}$ \\
& \texttt{TTIS}   &                           & $\textbf{93.08}_{\textbf{1.12}}$ & $\underline{0.995_{0.001}}$ & $\textbf{0.918}_{\textbf{0.017}}$
         &                            & $\textbf{94.25}_{\textbf{3.98}}$ & $\underline{0.993_{0.007}}$ & $\underline{0.938_{0.037}}$ \\

\midrule

\multirow{9}{*}{TCGA-NSCLC}
& None   & \multirow{3}{*}{ABMIL}    & $\textbf{92.58}_{\textbf{1.58}}$ & $\textbf{0.985}_{\textbf{0.012}}$ & $\textbf{0.926}_{\textbf{0.016}}$
         & \multirow{3}{*}{TransMIL} & $91.46_{3.87}$ & $\textbf{0.985}_{\textbf{0.012}}$ & $\underline{0.918_{0.033}}$ \\
& Random &                           & $\textbf{92.58}_{\textbf{1.58}}$ & $\textbf{0.985}_{\textbf{0.012}}$ & $\textbf{0.926}_{\textbf{0.016}}$
         &                          & $\textbf{91.97}_{\textbf{1.92}}$ & $\underline{0.984_{0.010}}$ & $\textbf{0.920}_{\textbf{0.019}}$ \\
& \texttt{TTIS}   &                           & $\textbf{92.58}_{\textbf{1.58}}$ & $\textbf{0.985}_{\textbf{0.013}}$ & $\textbf{0.926}_{\textbf{0.016}}$
         &                          & $\underline{91.95_{2.30}}$ & $0.983_{0.012}$ & $0.912_{0.014}$ \\
\cmidrule(lr){2-10}
& None   & \multirow{3}{*}{CLAM-SB}  & $\underline{92.99_{4.93}}$ & $\textbf{0.982}_{\textbf{0.015}}$ & $0.930_{0.049}$
         & \multirow{3}{*}{WiKG}     & $90.25_{2.10}$ & $0.972_{0.015}$ & $0.902_{0.021}$ \\
& Random &                           & $92.61_{4.55}$ & $\textbf{0.982}_{\textbf{0.016}}$ & $\textbf{0.939}_{\textbf{0.060}}$
         &                          & $\underline{90.91_{0.60}}$ & $\textbf{0.977}_\textbf{0.011}$ & $\underline{0.909_{0.006}}$ \\
& \texttt{TTIS}   &                           & $\textbf{93.06}_{\textbf{4.07}}$ & $\textbf{0.982}_{\textbf{0.015}}$ & $\underline{0.931_{0.041}}$
         &                          & $\textbf{91.75}_{\textbf{1.20}}$ & $\underline{0.976_{0.012}}$ & $\textbf{0.917}_{\textbf{0.012}}$ \\
\cmidrule(lr){2-10}
& None   & \multirow{3}{*}{S4MIL}    & $\underline{92.52_{1.53}}$ & $\textbf{0.983}_{\textbf{0.005}}$ & $\textbf{0.925}_{\textbf{0.015}}$
         & \multirow{3}{*}{R$^2$T-MIL} & $\underline{93.37_{4.28}}$ & $\textbf{0.985}_{\textbf{0.017}}$ & $\underline{0.934_{0.043}}$ \\
& Random &                           & $92.28_{1.05}$ & $\underline{0.982_{0.004}}$ & $\underline{0.923_{0.010}}$
         &                            & $\textbf{93.59}_{\textbf{4.33}}$ & $\textbf{0.985}_{\textbf{0.016}}$ & $\textbf{0.936}_{\textbf{0.043}}$ \\
& \texttt{TTIS}   &                           & $\textbf{92.55}_{\textbf{0.58}}$ & $\underline{0.982_{0.003}}$ & $\textbf{0.925}_{\textbf{0.006}}$
         &                            & $\underline{93.37_{4.28}}$ & $\textbf{0.985}_{\textbf{0.018}}$ & $\underline{0.934_{0.043}}$ \\

\bottomrule
\end{tabular}%
}
\end{table}

\section{Results}
\subsection{Main Results}
Table~\ref{tab:results-main} reports the classification and subtyping performance of the six MIL models under three conditions: (1) baseline (no instance selection), (2) with random sampling, and (3) with \texttt{TTIS}. 
Overall, \texttt{TTIS} consistently improves or preserves performance relative to baseline (processing all patches). On CAMELYON16, \texttt{TTIS} demonstrated substantial improvements over both baselines and random sampling. Compared to baselines, it achieved the best accuracy and F1 score in all models with gains of 0.52\%-3.62\% and 0.005-0.041, respectively.  Random sampling consistently underperformed both baselines and \texttt{TTIS}, indicating that naive instance selection degrades performance. 

Moreover, similar trends were observed across the four TCGA cohorts.  \texttt{TTIS} generally matched or improved baseline performance across the six MIL models. 
On TCGA-BRCA, \texttt{TTIS} outperformed baselines with respect to accuracy and F1 score in 5 out of 6 models, yielding notable improvements in accuracy by up to 2.09\% (ABMIL) and F1-score by up to 0.036 (WiKG). 
On TCGA-ESCA, \texttt{TTIS} yielded consistent gains for ABMIL and TransMIL in all evaluation metrics (e.g., 2.40\% and 1.19\% in accuracy and 0.025 and 0.012 in F1 score), and improvements in accuracy and F1 score for CLAM-SB, S4MIL, and R$^2$T-MIL, while maintaining competitive AUC. 
On TCGA-RCC and TCGA-NSCLC, \texttt{TTIS} largely preserved baseline performance, with modest gains for several models (e.g., TransMIL/S4MIL on TCGA-RCC and CLAM-SB/WiKG on TCGA-NSCLC) with negligible differences in AUC. 
Specifically, on TCGA-RCC, it matched or increased accuracy in 4 out of 6 models (e.g., TransMIL from 95.00\% to 95.77\% and S4MIL from 92.31\% to 93.08\%), and on TCGA-NSCLC, it matched or exceeded baseline accuracy across all six models, highlighted by a 1.50\% gain for WiKG. 
In these cohorts, overall performance differences between baseline, random sampling, and \texttt{TTIS} were minimal. These marginal improvements are likely attributable to inherent performance saturation, as evidenced by CLAM-SB on TCGA-RCC and ABMIL on TCGA-NSCLC, where all three experimental conditions obtained identical performance. 

\begin{table}[ht]
\centering
\caption{Ablation study on instance selection and ensemble, the number of clusters, and sampling ratio ($mean_{std}$).}
\label{tab:results-ablation}
\setlength{\tabcolsep}{3pt}
\renewcommand{\arraystretch}{1.05}
\resizebox{0.95\textwidth}{!}{%
\begin{tabular}{@{}c|c|ccc|ccc|ccc@{}}
\toprule
\multirow{2}{*}{Ablation} & \multirow{2}{*}{Settings} &  ACC & AUC & F1 & ACC & AUC & F1 & ACC & AUC & F1 \\
  \cmidrule(lr){3-5} \cmidrule(lr){6-8} \cmidrule(lr){9-11} 
 & & \multicolumn{3}{c|}{CAMELYON16} & \multicolumn{3}{c|}{TCGA-BRCA} & \multicolumn{3}{c}{TCGA-ESCA}  \\
\midrule
\cmidrule(lr){1-11}
\multirow[c]{5}{*}{\shortstack[c]{Instance\\Selection\\and\\Ensemble}}
 & Baseline         & $96.64_{1.18}$ & $0.994_{0.002}$ & $0.964_{0.013}$ & $89.85_{3.65}$ & $0.968_{0.021}$ & $0.825_{0.076}$ & $92.19_{4.62}$ & $0.970_{0.019}$ & $0.920_{0.047}$ \\
 & Class-guided     & $97.16_{0.44}$ & $0.993_{0.001}$ & $0.969_{0.005}$ & $91.94_{4.32}$ & $0.968_{0.019}$ & $0.854_{0.088}$ & $94.80_{3.97}$ & $0.979_{0.029}$ & $0.945_{0.036}$ \\
 & Centroid-based   & $97.16_{0.44}$ & $0.991_{0.002}$ & $0.969_{0.005}$ & $91.94_{4.32}$ & $0.968_{0.019}$ & $0.854_{0.088}$ & $95.39_{4.07}$ & $0.979_{0.028}$ & $0.951_{0.038}$ \\
 & Attention-based  & $97.16_{0.44}$ & $0.992_{0.002}$ & $0.969_{0.005}$ & $91.53_{3.91}$ & $0.968_{0.018}$ & $0.846_{0.080}$ & $95.39_{4.07}$ & $0.979_{0.029}$ & $0.951_{0.038}$ \\
 & \texttt{TTIS}    & $97.16_{0.44}$ & $0.992_{0.001}$ & $0.969_{0.005}$ & $91.94_{4.32}$ & $0.968_{0.019}$ & $0.854_{0.088}$ & $94.59_{3.19}$ & $0.972_{0.023}$ & $0.945_{0.032}$ \\
\cmidrule(lr){1-11}
\multirow[c]{5}{*}{\shortstack[c]{Number\\of\\Clusters}}
 & $K = 0$ & $96.64_{1.18}$ & $0.994_{0.002}$ & $0.964_{0.013}$ & $89.85_{3.65}$ & $0.968_{0.021}$ & $0.825_{0.076}$ & $92.19_{4.62}$ & $0.970_{0.019}$ & $0.920_{0.047}$ \\
 & $K = 2$    & $96.64_{0.45}$ & $0.993_{0.002}$ & $0.966_{0.005}$ & $90.70_{3.44}$ & $0.968_{0.019}$ & $0.841_{0.065}$ & $92.79_{3.67}$ & $0.971_{0.020}$ & $0.927_{0.037}$ \\
 & $K = 4$    & $97.16_{0.44}$ & $0.992_{0.002}$ & $0.967_{0.005}$ & $90.26_{4.15}$ & $0.969_{0.019}$ & $0.831_{0.083}$ & $91.59_{4.60}$ & $0.971_{0.018}$ & $0.914_{0.047}$ \\
 & $K = 8$    & $97.16_{0.45}$ & $0.992_{0.003}$ & $0.969_{0.005}$ & $90.68_{4.35}$ & $0.968_{0.019}$ & $0.831_{0.086}$ & $91.59_{4.60}$ & $0.970_{0.021}$ & $0.914_{0.047}$ \\
 & $K = 16$   & $97.16_{0.45}$ & $0.994_{0.002}$ & $0.969_{0.005}$ & $90.68_{4.35}$ & $0.969_{0.019}$ & $0.836_{0.086}$ & $92.20_{3.80}$ & $0.969_{0.022}$ & $0.921_{0.039}$ \\
\cmidrule(lr){1-11}
\multirow[c]{9}{*}{\shortstack[c]{Sampling\\Ratio}}
 & $r = 100\%$    & $96.64_{0.01}$ & $0.994_{0.002}$ & $0.964_{0.013}$ & $89.85_{0.04}$ & $0.968_{0.021}$ & $0.825_{0.076}$ & $92.19_{0.05}$ & $0.970_{0.019}$ & $0.920_{0.047}$ \\
 & $r = 90\%$     & $96.64_{0.01}$ & $0.994_{0.001}$ & $0.964_{0.013}$ & $90.26_{0.04}$ & $0.968_{0.020}$ & $0.831_{0.083}$ & $92.79_{0.04}$ & $0.970_{0.018}$ & $0.927_{0.037}$ \\
 & $r = 75\%$     & $96.90_{0.01}$ & $0.993_{0.001}$ & $0.967_{0.009}$ & $90.68_{0.04}$ & $0.968_{0.021}$ & $0.836_{0.086}$ & $92.79_{0.04}$ & $0.971_{0.020}$ & $0.927_{0.037}$ \\
 & $r = 50\%$     & $97.16_{0.00}$ & $0.993_{0.001}$ & $0.969_{0.005}$ & $91.11_{0.04}$ & $0.969_{0.020}$ & $0.842_{0.077}$ & $93.40_{0.03}$ & $0.972_{0.021}$ & $0.933_{0.028}$ \\
 & $r = 40\%$     & $97.16_{0.00}$ & $0.993_{0.001}$ & $0.969_{0.005}$ & $91.11_{0.04}$ & $0.969_{0.020}$ & $0.842_{0.077}$ & $93.99_{0.03}$ & $0.973_{0.021}$ & $0.939_{0.029}$ \\
 & $r = 30\%$     & $97.16_{0.00}$ & $0.992_{0.001}$ & $0.969_{0.005}$ & $91.94_{0.04}$ & $0.968_{0.019}$ & $0.854_{0.088}$ & $94.59_{0.03}$ & $0.972_{0.023}$ & $0.945_{0.032}$ \\
 & $r = 20\%$     & $97.42_{0.00}$ & $0.991_{0.003}$ & $0.972_{0.005}$ & $91.96_{0.03}$ & $0.969_{0.019}$ & $0.851_{0.070}$ & $94.59_{0.03}$ & $0.970_{0.025}$ & $0.945_{0.032}$ \\
 & $r = 10\%$     & $97.16_{0.00}$ & $0.988_{0.005}$ & $0.970_{0.005}$ & $92.40_{0.03}$ & $0.968_{0.015}$ & $0.860_{0.055}$ & $93.98_{0.04}$ & $0.969_{0.027}$ & $0.939_{0.043}$ \\
\bottomrule
 &  & \multicolumn{3}{c|}{TCGA-RCC} & \multicolumn{3}{c|}{TCGA-NSCLC} & \multicolumn{3}{c}{-}  \\
 \midrule
\cmidrule(lr){1-11}
\multirow[c]{5}{*}{\shortstack[c]{Instance\\Selection\\and\\Ensemble}}
 & Baseline         & $96.16_{2.39}$ & $0.995_{0.005}$ & $0.956_{0.022}$ & $92.58_{1.58}$ & $0.985_{0.012}$ & $0.926_{0.016}$ & - & - & -  \\
 & Class-guided     & $93.87_{1.33}$ & $0.991_{0.002}$ & $0.935_{0.017}$ & $92.58_{1.58}$ & $0.984_{0.013}$ & $0.926_{0.016}$ & - & - & -  \\
 & Centroid-based   & $93.48_{1.76}$ & $0.992_{0.001}$ & $0.931_{0.019}$ & $91.81_{2.09}$ & $0.985_{0.013}$ & $0.918_{0.021}$ & - & - & -  \\
 & Attention-based  & $93.87_{1.33}$ & $0.990_{0.003}$ & $0.935_{0.017}$ & $92.20_{2.23}$ & $0.984_{0.013}$ & $0.922_{0.022}$ & - & - & - \\
 & \texttt{TTIS}    & $95.78_{2.89}$ & $0.995_{0.004}$ & $0.951_{0.030}$ & $92.58_{1.58}$ & $0.985_{0.013}$ & $0.926_{0.016}$ & - & - & -  \\
 \cmidrule(lr){1-11}
 \multirow[c]{5}{*}{\shortstack[c]{Number\\of\\Clusters}}
 & $K = 0$ & $96.16_{2.39}$ & $0.995_{0.005}$ & $0.956_{0.022}$ & $92.58_{1.58}$ & $0.985_{0.012}$ & $0.926_{0.016}$ & - & - & -  \\
 & $K = 2$    & $95.40_{3.44}$ & $0.995_{0.004}$ & $0.948_{0.036}$ & $92.58_{1.58}$ & $0.985_{0.012}$ & $0.926_{0.016}$ & - & - & -  \\
 & $K = 4$    & $95.78_{2.89}$ & $0.995_{0.004}$ & $0.951_{0.030}$ & $92.58_{1.58}$ & $0.983_{0.012}$ & $0.926_{0.016}$ & - & - & -  \\
 & $K = 8$    & $95.78_{2.89}$ & $0.994_{0.005}$ & $0.949_{0.030}$ & $92.58_{1.58}$ & $0.984_{0.013}$ & $0.926_{0.016}$ & - & - & -  \\
 & $K = 16$   & $95.78_{2.89}$ & $0.994_{0.005}$ & $0.949_{0.030}$ & $92.58_{1.58}$ & $0.984_{0.013}$ & $0.926_{0.016}$ & - & - & -  \\
\cmidrule(lr){1-11}
\multirow[c]{9}{*}{\shortstack[c]{Sampling\\Ratio}}
 & $r = 100\%$    & $96.16_{0.02}$ & $0.995_{0.005}$ & $0.956_{0.022}$ & $92.58_{0.02}$ & $0.985_{0.012}$ & $0.926_{0.016}$ & - & - & -  \\
 & $r = 90\%$     & $95.78_{0.03}$ & $0.994_{0.005}$ & $0.949_{0.030}$ & $92.58_{0.02}$ & $0.985_{0.011}$ & $0.926_{0.016}$ & - & - & -  \\
 & $r = 75\%$     & $95.78_{0.03}$ & $0.995_{0.004}$ & $0.949_{0.030}$ & $92.58_{0.02}$ & $0.984_{0.012}$ & $0.926_{0.016}$ & - & - & -  \\
 & $r = 50\%$     & $95.78_{0.03}$ & $0.995_{0.004}$ & $0.951_{0.030}$ & $92.58_{0.02}$ & $0.984_{0.013}$ & $0.926_{0.016}$ & - & - & -  \\
 & $r = 40\%$     & $95.78_{0.03}$ & $0.995_{0.004}$ & $0.951_{0.030}$ & $92.58_{0.02}$ & $0.984_{0.013}$ & $0.926_{0.016}$ & - & - & -  \\
 & $r = 30\%$     & $95.78_{0.03}$ & $0.995_{0.004}$ & $0.951_{0.030}$ & $92.58_{0.02}$ & $0.985_{0.013}$ & $0.926_{0.016}$ & - & - & -  \\
 & $r = 20\%$     & $95.78_{0.03}$ & $0.995_{0.005}$ & $0.951_{0.030}$ & $92.20_{0.02}$ & $0.984_{0.013}$ & $0.922_{0.022}$ & - & - & -  \\
 & $r = 10\%$     & $95.78_{0.03}$ & $0.995_{0.005}$ & $0.951_{0.030}$ & $92.20_{0.02}$ & $0.984_{0.015}$ & $0.922_{0.022}$ & - & - & -  \\
\bottomrule
\end{tabular}%
}
\end{table}

\subsection{Ablation Studies}
To systematically evaluate \texttt{TTIS}, we conducted three ablation experiments: (1) instance selection and ensemble, (2) number of hierarchical clusters $K$, and (3) sampling ratio $r$.
Table~\ref{tab:results-ablation} shows the results of ablation experiments on these three components using ABMIL, which frequently obtained the best overall performance among the six MIL models.

For instance selection and ensemble, we analyzed the individual performance of each initialization method. Each individual initialization method generally improved performance compared to the baseline (i.e., no instance selection), with the exception of TCGA-RCC. Notably, the proposed multi-view test-time ensemble, combining all three initialization methods, matched or provided modest improvements over any single method, indicating that the different strategies capture complementary sets of instances, and their integration provides more robust WSI representations.

To investigate the effectiveness of hierarchical clustering, we varied $K \in \{2,4,8,16\}$ and compared them against a baseline that does not conduct clustering ($K=0$). Except for TCGA-RCC, hierarchical clustering provided consistent improvements over the baseline and remained remarkably stable with varying values of $K$. This indicates that \texttt{TTIS} is highly robust to the specific choice of cluster count, and hierarchical clustering captures spatial and semantic heterogeneity.

To understand the trade-off between efficiency and accuracy, we varied the sampling ratio $r$ from 100\% to 10\%. Overall, \texttt{TTIS} exhibited remarkable robustness under substantial downsampling. Notably, \texttt{TTIS} with $r<$100\% often outperformed the baseline, demonstrating that processing all patches is sub-optimal. 
These findings empirically validate the core premise of \texttt{TTIS}: WSIs contain substantial quantities of redundant and uninformative patches. Removing these redundant and non-diagnostic patches at inference suppresses noise that would otherwise confound the MIL aggregation, thereby improving overall predictive performance. 

\section{Conclusion}
We propose $\texttt{TTIS}$, a training-free, plug-and-play test-time instance selection framework for WSI analysis. Since $\texttt{TTIS}$ requires no retraining or architecture modifications to pre-trained MIL models, it guarantees seamless integration into existing computational pathology pipelines. 
By employing hierarchical spatio-semantic clustering followed by $\texttt{FPS}$, $\texttt{TTIS}$ constructs a compact yet representative patch subset directly at inference time. To further improve robustness, $\texttt{TTIS}$ employs a multi-view ensemble strategy that combines patch subsets derived from complementary initialization criteria: centroid-based, attention-based, and class-guided, each capturing a distinct facet of tissue representation. Extensive experiments across five public datasets demonstrate that $\texttt{TTIS}$ consistently improves or matches the performance of baseline MIL models. 

\bibliographystyle{splncs04}
\bibliography{paper}
\end{document}